\documentclass[twocolumn,showpacs,aps,prl,superscriptaddress,floatfix]{revtex4}

\usepackage{graphicx}
\usepackage{dcolumn}
\usepackage{amsmath}
\usepackage{epsfig}
\usepackage{multirow}

\def\ifb{\;\mbox{fb}^{-1}}

\def\numBB{$211\pm 2$ million}

\def\GeV{\;\mbox{GeV}}

\def\GeVcc{\;\mbox{GeV}/c^2}
\def\MeV{\;\mbox{MeV}}
\def\MeVcc{\;\mbox{MeV}/c^2}

\def\thd  {\theta_D}
\def\de   {\Delta E^*}

\def\mes  {M_{\mbox{\scriptsize ES}} }
\def\bkg  {B \to K^{*}\gamma}

\def\brg  {B \to \rho\gamma}

\def\brpg {B^+ \to \rho^+\gamma}
\def\brzg {B^0 \to \rho^0\gamma}
\def\bomg {B^0 \to \omega\gamma}

\def\bsg  {b\to s\gamma}

\def\rhoz {\rho^0}
\def\rhop {\rho^+}

\def\bdg    {\ensuremath {b \to d \gamma}}
\def\bkg    {\ensuremath {\B \to \Kstar \gamma}}

\def\Kz    {\ensuremath{K^{0}}\xspace}

\def\de        {\ensuremath {\Delta E^{*}}}

\def\avbr      {\ensuremath{\overline{\BR}[B\rightarrow(\rho/\omega)\gamma]}}

\newcommand{\BaBarYear}       {04}
\newcommand{\BaBarNumber}     {35}
\newcommand{\SLACPubNumber} {10608}

\RequirePackage{xspace}

\usepackage{relsize}
\def\babar{\mbox{\slshape B\kern-0.1em{\smaller A}\kern-0.1em
    B\kern-0.1em{\smaller A\kern-0.2em R}}}

\def\en         {\ensuremath{e^-}\xspace} 
\def\ep         {\ensuremath{e^+}\xspace}
 
\def\epem       {\ensuremath{e^+e^-}\xspace}

\def\s     {\ensuremath{s}\xspace}

\def\b     {\ensuremath{b}\xspace}

\def\piz   {\ensuremath{\pi^0}\xspace}

\def\pip   {\ensuremath{\pi^+}\xspace}
\def\pim   {\ensuremath{\pi^-}\xspace}

\def\pipm  {\ensuremath{\pi^\pm}\xspace}

\def\Kbar  {\kern 0.2em\overline{\kern -0.2em K}{}\xspace}

\def\Kz    {\ensuremath{K^0}\xspace}
\def\Kzb   {\ensuremath{\Kbar^0}\xspace}
\def\KzKzb {\ensuremath{\Kz \kern -0.16em \Kzb}\xspace}
\def\Kp    {\ensuremath{K^+}\xspace}
\def\Km    {\ensuremath{K^-}\xspace}
\def\Kpm   {\ensuremath{K^\pm}\xspace}

\def\KpKm  {\ensuremath{\Kp \kern -0.16em \Km}\xspace}
\def\KS    {\ensuremath{K^0_{\scriptscriptstyle S}}\xspace}

\def\Kstar   {\ensuremath{K^*}\xspace}

\def\Dbar    {\kern 0.2em\overline{\kern -0.2em D}{}\xspace}

\def\Dz      {\ensuremath{D^0}\xspace}
\def\Dzb     {\ensuremath{\Dbar^0}\xspace}
\def\DzDzb   {\ensuremath{\Dz {\kern -0.16em \Dzb}}\xspace}
\def\Dp      {\ensuremath{D^+}\xspace}
\def\Dm      {\ensuremath{D^-}\xspace}

\def\DpDm    {\ensuremath{\Dp {\kern -0.16em \Dm}}\xspace}

\def\B       {\ensuremath{B}\xspace}
\def\Bbar    {\kern 0.18em\overline{\kern -0.18em B}{}\xspace}

\def\BB      {\ensuremath{B\Bbar}\xspace} 
\def\Bz      {\ensuremath{B^0}\xspace}
\def\Bzb     {\ensuremath{\Bbar^0}\xspace}
\def\BzBzb   {\ensuremath{\Bz {\kern -0.16em \Bzb}}\xspace}
\def\Bu      {\ensuremath{B^+}\xspace}
\def\Bub     {\ensuremath{B^-}\xspace}
\def\Bp      {\ensuremath{\Bu}\xspace}

\def\BpBm    {\ensuremath{\Bu {\kern -0.16em \Bub}}\xspace}

\def\BorBbar    {\kern 0.18em\optbar{\kern -0.18em B}{}\xspace}
\def\DorDbar    {\kern 0.18em\optbar{\kern -0.18em D}{}\xspace}
\def\KorKbar    {\kern 0.18em\optbar{\kern -0.18em K}{}\xspace}

\mathchardef\Upsilon="7107
\def\Y#1S{\ensuremath{\Upsilon{(#1S)}}\xspace}

\def\FourS {\Y4S}

\mathchardef\Deltares="7101
\mathchardef\Xi="7104
\mathchardef\Lambda="7103
\mathchardef\Sigma="7106
\mathchardef\Omega="710A

\def\Deltabar{\kern 0.25em\overline{\kern -0.25em \Deltares}{}\xspace}
\def\Lbar{\kern 0.2em\overline{\kern -0.2em\Lambda\kern 0.05em}\kern-0.05em{}\xspace}
\def\Sigbar{\kern 0.2em\overline{\kern -0.2em \Sigma}{}\xspace}
\def\Xibar{\kern 0.2em\overline{\kern -0.2em \Xi}{}\xspace}
\def\Obar{\kern 0.2em\overline{\kern -0.2em \Omega}{}\xspace}
\def\Nbar{\kern 0.2em\overline{\kern -0.2em N}{}\xspace}
\def\Xb{\kern 0.2em\overline{\kern -0.2em X}{}\xspace}

\def\BR         {{\ensuremath{\cal B}\xspace}}

\def\mes        {\mbox{$m_{\rm ES}$}\xspace}

\def\DeltaE     {\mbox{$\Delta E$}\xspace}

\newcommand{\tev}{\ensuremath{\mathrm{\,Te\kern -0.1em V}}\xspace}
\newcommand{\gev}{\ensuremath{\mathrm{\,Ge\kern -0.1em V}}\xspace}
\newcommand{\mev}{\ensuremath{\mathrm{\,Me\kern -0.1em V}}\xspace}
\newcommand{\kev}{\ensuremath{\mathrm{\,ke\kern -0.1em V}}\xspace}
\newcommand{\ev}{\ensuremath{\mathrm{\,e\kern -0.1em V}}\xspace}
\newcommand{\gevc}{\ensuremath{{\mathrm{\,Ge\kern -0.1em V\!/}c}}\xspace}
\newcommand{\mevc}{\ensuremath{{\mathrm{\,Me\kern -0.1em V\!/}c}}\xspace}
\newcommand{\gevcc}{\ensuremath{{\mathrm{\,Ge\kern -0.1em V\!/}c^2}}\xspace}
\newcommand{\mevcc}{\ensuremath{{\mathrm{\,Me\kern -0.1em V\!/}c^2}}\xspace}

\def\mus  {\ensuremath{\rm \,\mus}\xspace}

\def\mus        {\ensuremath{\,\mu{\rm s}}\xspace}

\def\to                 {\ensuremath{\rightarrow}\xspace}

\def\pep2{PEP-II}

\def\gsim{{~\raise.15em\hbox{$>$}\kern-.85em
          \lower.35em\hbox{$\sim$}~}\xspace}
\def\lsim{{~\raise.15em\hbox{$<$}\kern-.85em
          \lower.35em\hbox{$\sim$}~}\xspace}

\def\Vtd  {\ensuremath{|V_{td}|}\xspace}

\def\Vts  {\ensuremath{|V_{ts}|}\xspace}

\def\Vcb  {\ensuremath{|V_{cb}|}\xspace}

\xspace

\newcommand{\jprlBase}       {Phys.\ Rev.\ Lett.\xspace}

\newcommand{\jplBase}        {Phys.\ Lett.\xspace}

\newcommand{\npBase}         {Nucl.\ Phys.\xspace}

\newcommand{\arnps}     [1]  {{Ann.\ Rev.\ Nucl.\ Part.\ Sci.\ {\bf #1}}}

\newcommand{\cpc}       [1]  {{Comput.\ Phys.\ Commun.\ {\bf #1}}}

\newcommand{\npb}       [1]  {\npBase\ B~{\bf #1}}

\newcommand{\plb}       [1]  {\jplBase\ B~{\bf #1}}

\newcommand{\jprl}      [1]  {\jprlBase\ {\bf #1}}

\newcommand{\hepex}     [1]  {hep-ex/{#1}}

\def\jetset74   {\mbox{\tt Jetset \hspace{-0.5em}7.\hspace{-0.2em}4}\xspace}

\begin{document}

{\pagestyle{empty}
\begin{flushleft}
\babar-PUB-\BaBarYear/\BaBarNumber \\
SLAC-PUB-\SLACPubNumber \\
\end{flushleft}
\begin{flushright}
\end{flushright}
}

\title{Search for Radiative Penguin Decays $\brpg$, $\brzg$, and $\bomg$}

\date{\today}

\begin{abstract}
A search for the decays
$B\to\rho(770)\gamma$ and $\Bz\to\omega(782)\gamma$
is performed on a sample of 211 million 
$\FourS\rightarrow\BB$ events collected by the $\babar$ detector at
the \pep2 asymmetric-energy
$\epem$ storage ring. No evidence for the decays is seen. 
We set the following limits on 
the individual branching fractions $\BR(\brpg)<1.8\times 10^{-6}$, $\BR(\brzg)<0.4\times 10^{-6}$, and
$\BR(\bomg)<1.0\times 10^{-6}$ at the 90\% confidence level (C.L.). We use the quark 
model to 
limit the combined branching fraction $\avbr<1.2\times 10^{-6}$, from which we 
determine a constraint on the ratio of CKM matrix elements $\Vtd/\Vts$.  
\end{abstract}


\author{B.~Aubert}
\author{R.~Barate}
\author{D.~Boutigny}
\author{F.~Couderc}
\author{J.-M.~Gaillard}
\author{A.~Hicheur}
\author{Y.~Karyotakis}
\author{J.~P.~Lees}
\author{V.~Tisserand}
\author{A.~Zghiche}
\affiliation{Laboratoire de Physique des Particules, F-74941 Annecy-le-Vieux, France }
% ...
\collaboration{The \babar\ Collaboration}
\noaffiliation

\maketitle

Within the Standard Model (SM), the decays 
$\B\rightarrow\rho\gamma$ and $\bomg$ proceed
primarily through a $\bdg$ electromagnetic penguin
process that contains a top quark within the loop \cite{review}. 
The rates for 
$\brpg$, $\brzg$, and $\bomg$ \cite{ccmodes} are related by the spectator-quark
 model, and we define the average branching fraction \cite{ali2004},
       $\avbr  = \frac{1}{2} \left\{ \BR(\brpg) + \frac{\tau_{\Bp}}{\tau_{\Bz}} [ \BR(\brzg) + \BR(\bomg)]\right\}$, 
where $\frac{\tau_{\Bp}}{\tau_{\Bz}}$ is the ratio of \B-meson lifetimes.
Recent calculations of $\avbr$ in the SM indicate a range of
$(0.9-1.8)\times 10^{-6}$ \cite{SM,ali2004}. There may also be contributions 
resulting from physics beyond
the SM \cite{hewett}.
The ratio between the branching fractions for 
$\B\rightarrow(\rho/\omega)\gamma$ and $\bkg$ is related in the SM to 
the ratio 
of Cabibbo-Kobayashi-Maskawa (CKM) matrix elements $\Vtd/\Vts$ \cite{alivtdvtstheory,ali2004}.
Previous searches by $\babar$~\cite{oldbabar} 
and CLEO \cite{cleobellerg} have found no evidence for 
 $\B\rightarrow(\rho/\omega)\gamma$ decays.

We search for $\brg$ and $\bomg$ decays in a data sample containing 
\numBB\ $\FourS\rightarrow\BB$ decays, collected by the \babar\ 
detector~\cite{ref:detector} at the \pep2 asymmetric-energy $\epem$ storage 
ring. The data correspond to an integrated luminosity 
of 191$\ifb$. 

The decay $\brg$ is reconstructed with $\rho^0\to\pip\pim$
and $\rho^+\to\pip\piz$,
while $\bomg$ is reconstructed with $\omega\to\pip\pim\piz$.
Background comes primarily from
$\ep\en \to q\bar{q}$ continuum events, where $q=u,d,s,c$, in which  
a high-energy photon is produced through
$\piz/\eta\to\gamma\gamma$ decays or via initial-state radiation (ISR).
There are also significant $\BB$ backgrounds:
$\bkg$, $\Kstar\rightarrow K\pi$, where 
a $\Kpm$ is misidentified as a $\pipm$;
$\B\rightarrow(\rho/\omega)\piz$ and $\B\rightarrow(\rho/\omega)\eta$,  where 
a high-energy photon  comes from the $\piz$ or $\eta$ decay; and
combinatorial background, mostly 
from high multiplicity $\b\rightarrow\s\gamma$ decays.  

We select $\pipm$ candidates from tracks with
a momentum transverse to the beam direction 
greater than $100~\mevc$. The $\pipm$ selection
algorithm combines measurements of energy loss in the
tracking system
with any associated Cherenkov photons measured by the ring imaging Cherenkov 
detector. The 
algorithm is optimized to reduce backgrounds from $\Kpm$ produced 
in $\bsg$ processes \cite{oldbabar}. 

Neutral pion
candidates are identified as pairs of neutral energy-deposits reconstructed
in the CsI crystal calorimeter, each with an energy greater than $50\mev$ 
in the laboratory frame.
For $\Bz\rightarrow\omega\gamma$ $(\brpg)$ decays, the invariant mass of the pair is required to satisfy $110 < m_{\gamma\gamma} < 150\MeVcc$
$(117 < m_{\gamma\gamma} < 145\MeVcc)$. To reduce combinatorial 
background, we require the cosine of the 
opening angle between the daughter photons in the laboratory frame 
be greater than 0.6; this selection retains 98\% of $\piz$ from 
signal decays. 

A $\rho^0$ candidate is
reconstructed by
selecting two identified pions that have opposite charge and
a common vertex.
We obtain $\rho^+$ candidates by pairing $\piz$
candidates with an identified $\pip$.
The $\omega$ candidates are reconstructed by combining
a $\piz$ candidate with pairs of 
oppositely charged pion candidates that originate from a common vertex; 
the charged pion pair must be 
consistent with originating from the interaction region to suppress 
$\KS$ decays.  We select $\rho$ $(\omega)$ candidates with an invariant mass 
satisfying $630 < m_{\pi\pi} < 940\MeVcc$ ($764 < m_{\pip\pim\piz} < 795\MeVcc$).

The high-energy photon from the signal $B$ decay is identified as a neutral
energy-deposit in the calorimeter. 
We require that the deposit
meets a number of criteria designed to eliminate background 
from charged particles and hadronic showers \cite{babarksg}. We 
veto photons from $\piz(\eta)$ decay by requiring that the 
invariant mass of the candidate combined with any other photon 
of laboratory energy greater than $30~(250)\MeV$ not to be within the 
range $105$ to $155~\MeVcc$ ($500$ to $590~\MeVcc$).

The photon and $\rho/\omega$ candidates are
combined to form the $B$-meson candidates.
We define $\de \equiv E^*_{B}-E_{\rm beam}^*$,
where 
$E^*_B$ is the center-of-mass (CM) energy
of the $B$-meson candidate and $E_{\rm beam}^*$ is the CM beam energy. 
The $\de$ distribution of Monte Carlo (MC) simulated signal events is 
centered at zero, with a resolution of about $50\mev$. 
We also define the beam-energy-substituted mass
$\mes \equiv
\sqrt{ E^{*2}_{\rm beam}-{\mathrm p}_{B}^{*2}}$,
where ${\mathrm p}_B^*$ is the CM momentum of the $B$ candidate. Signal MC 
events peak in $\mes$ at the mass of the $B$-meson, $m_B$, with a resolution 
of $3~\mevcc$. The distribution of continuum
and combinatorial $\BB$ background peaks in neither $\mes$ nor $\DeltaE^{*}$; 
the background distributions of $\bkg$, $\B\rightarrow(\rho/\omega)\piz$ and 
$\B\rightarrow(\rho/\omega)\eta$ peak at $m_B$ in $\mes$   
and between -190~\mev\ and -60~\mev\ in $\DeltaE^{*}$. 
We consider candidates in the ranges $-0.3 < \de <0.3 \GeV$ and
$5.20 < \mes<  5.29\GeVcc$
to incorporate sidebands that allow the combinatorial background yields
to be extracted from 
a fit to the data.

Several variables that distinguish between signal 
and continuum events are combined in a neural network \cite{snns}. The input variables depend mainly on the rest of the 
event (ROE), defined to be all 
charged tracks and neutral energy deposits in the calorimeter 
not used to reconstruct the $B$ candidate.
To reject ISR events, we compute the ratio of second-to-zeroth order 
Fox-Wolfram moments~\cite{fox} for the ROE and the $\rho/\omega$ candidate, 
in the frame recoiling against the photon momentum. To discriminate between the jet-like continuum background 
and the more spherically-symmetric signal events, we compute the angle 
between the photon and the thrust axis of the ROE in the CM frame 
and the moments 
$L_{i} \equiv \sum_{j} 
p^{*}_{j}\cdot|\cos{\theta^{*}_{j}}|^{i}/\sum_{j} p^{*}_{j}$,
where
$p^{*}_j$ and $\theta^{*}_{j}$ are the momentum and angle with respect 
to an axis, respectively, for each particle $j$ in the ROE.
We use $L_{1}$, $L_{2}$, and $L_{3}$ with respect to the thrust axis of the
ROE, as well as $L_{1}$ with respect to the photon direction.
Differences in lepton and kaon production between background  and \B decays
are exploited by including \babar\ flavor tagging  
variables~\cite{babartag} as well as the maximum CM momentum and number  
of \Kpm\ and \KS\ in the ROE.
For the $\Bz\rightarrow(\rhoz/\omega)\gamma$ modes, we also use the 
separation along the beam axis of the $B$-meson candidate and ROE 
vertices; to remove poorly reconstructed events we require the 
separation be less than 4~mm.
A separate neural network is trained for each mode.
We make a loose selection on the
output of the neural network, $\mathcal{N}$, that retains around 80\% of
the signal events. 

To suppress background, we combine a number of signal-decay variables in a Fisher  
discriminant \cite{fisher}, $\mathcal{F}$, separately for each mode. 
We calculate the $B$-meson production angle $\theta_B^*$,
the $\rho/\omega$ helicity angle 
$\theta_H$, which is defined with respect to the normal of the 
decay plane for $\omega$ candidates, and the $\omega$ 
Dalitz angle $\thd$ \cite{oldbabar}. To reject 
$\B\rightarrow\rho(\piz/\eta)$ and 
$\B\rightarrow\omega(\piz/\eta)$ events in the $\brpg$ and $\bomg$
$(\brzg)$ selection, we require
$|\cos\theta_{H}|< 0.70~(0.75)$.  

After applying the $\mathcal{N}$ and $|\cos\theta_H|$ criteria the expected 
average candidate multiplicity in signal events is 1.15, 1.03 and 1.14 for 
$\brpg$, $\brzg$ and $\bomg$, respectively; in events with multiple candidates
the one with the smallest value of $|\DeltaE^{*}|$ is retained. 

The signal yield is determined from an extended 
maximum likelihood fit to the selected data. We fit the 
four-dimensional distribution of $\mes$, $\DeltaE^{*}$, 
$\mathcal{F}$ and $\mathcal{N}$.
For the $\brg$ fits, five event hypotheses are considered: 
signal, continuum background, combinatorial $B$-background, 
peaking $\B\rightarrow\rho(\piz/\eta)$ 
background and peaking $\bkg$ background. For the $\bomg$ fit we consider 
only signal, continuum background, and peaking $\B\rightarrow\omega(\piz/\eta)$
background.
The correlations among the observables are small; therefore, we assume that the
 probability density function (PDF) $\mathcal{P}(\vec{x_j};
\vec{\alpha_{i}})$ for each hypothesis 
is the product of individual PDFs for the variables 
$\vec{x}_{j}=\{\mes,\DeltaE^{*},\mathcal{F},\mathcal{N}\}$ 
given the set of parameters $\vec{\alpha}_{i}$. The likelihood 
function is a product over
all $N_k$ candidate events of the sum of the PDFs,
 \begin{displaymath}
 {\cal L}_{k}=\exp{\left(-\sum_{i=1}^{N_{\mathrm{hyp}}} n_{i}\right)}\cdot\left[\prod_{j = 1}^{N_k}\left(\sum_{i=1}^{N_{\mathrm{hyp}}} n_i{\cal P}_{i}(\vec{x}_j;\ \vec{\alpha}_i)\right)\right]\; ,
 \end{displaymath}  
where $n_i$ is the yield of each hypothesis, $k$ is the $\B\rightarrow(\rho/\omega)\gamma$ mode, and $N_{\mathrm{hyp}}=5(3)$ for $\B\to\rho\gamma$ ($\B\to\omega\gamma$).

\begin{table*}[th]
\caption{\label{tab:results} The signal yield $(n_{\mathrm{sig}})$,
continuum background yield ($n_{\mathrm{cont}}$),
peaking background ($n_{\mathrm{peak}}$),
significance in standard deviations $\sigma$, efficiency $(\epsilon)$, and
branching fraction $(\mathcal{B})$ central value and upper limit at the 90\% C.L for each mode.
The results of the combined fit are shown in the bottom row where $n_{\mathrm{sig}}$ is equal to
$n_{\mathrm{eff}}$, which is described in the text.
When two errors are
quoted, the first is statistical and the second is systematic.}
 \begin{tabular}{l l c l c r l c}
\hline
\hline
Mode      &    \multicolumn{1}{c}{$n_{\mathrm{sig}}$}       & \multicolumn{1}{c}{$n_{\mathrm{cont}}$} &
\multicolumn{1}{c}{$n_{\mathrm{peak}}$} & \multicolumn{1}{l}{Significance $(\sigma)$} & \multicolumn{1}{c}{$\epsilon (\%)$} &
\multicolumn{1}{c}{$\mathcal{\B} (10^{-6})$} &  \multicolumn{1}{c}{$\mathcal{\B}(10^{-6})$ 90\% C.L.}\\
\hline
$\brpg$   &$\,26^{\,+15\,+2}_{\,-14\,-2}$   &$6850\pm90$   & $18\pm4$           & 1.9  &  $13.2\pm 1.4\ $  &  $0.9\ ^{+\
0.6}_{-\ 0.5}\,\pm 0.1$    &  $<1.8$ \\
$\brzg$   &$0.3^{+7.2+1.7}_{-5.4-1.6}$&$4269\pm73$   & $18\pm7$           & 0.0   &  $15.8\pm 1.9\ $  &   $0.0\pm 0.2
\pm0.1$   &  $<0.4$ \\
$\bomg$   &$8.3^{+5.7+1.3}_{-4.5-1.9}$&$1378\pm37$   & $2.6^{+0.8}_{-1.2}$& 1.5 &  $8.6\pm 0.9\ $   &  $0.5\pm0.3\pm0.1$   &
$<1.0$\\
\hline
Combined  & \multicolumn{1}{c}{$269^{+126+40}_{-120-45}$}    & \multicolumn{1}{c}{---} & \multicolumn{1}{c}{---} &  2.1 &
\multicolumn{1}{c}{---} &  $0.6\pm0.3\pm0.1$     & $<1.2$\\
\hline
\hline
\end{tabular}
\end{table*}

The $\mes$ and $\DeltaE^{*}$ PDFs are parameterized by a Crystal Ball 
function \cite{CryBall} for both the signal and peaking background. The 
parameterization is determined from signal MC samples, except the mean of 
the $\de$ distribution, which is offset by the observed difference 
between data and MC samples of $\bkg$ decays.
The continuum background $\mes$ and $\DeltaE^{*}$ distributions are 
parameterized by an ARGUS threshold function \cite{Argus} 
and a second-order polynomial, respectively. The combinatorial 
\B\ background is described by a smoothed distribution \cite{KEYS} determined from MC events in both 
$\mes$ and $\DeltaE^{*}$. The distribution of $\mathcal{N}$ for signal and 
\BB background is
parameterized by a Crystal Ball function. 
The  $\mathcal{N}$ distribution for continuum is determined from 
sideband data, and a histogram is used as the PDF. 
The distribution of $\mathcal{F}$ is
parameterized by smoothed histograms of sideband data for the continuum 
background and MC events for all other hypotheses. 

The fit to the data determines the shape parameters of the continuum background $\mes$
and $\de$ PDFs, as well as the signal, continuum background 
and combinatorial $\BB$ background yields. All other parameters are fixed, including 
the peaking $\BB$ background yields. A combined fit is also performed relating 
the modes using the definition of \avbr\ to determine an effective yield 
($n_{\mathrm{eff}}$) assuming $n(\brpg)=n_{\mathrm{eff}}\cdot\epsilon(\brpg)$ and 
$n(\Bz\rightarrow(\rhoz/\omega)\gamma)=\frac{1}{2}\frac{\tau_{\Bz}}{\tau_{\Bp}}n_{\mathrm{eff}}\cdot\epsilon(\Bz\rightarrow(\rhoz/\omega)\gamma)$ where $n$
and $\epsilon$
are the  yields and reconstruction efficiencies of 
each mode; the efficiencies include the daughter branching fractions. We take $\frac{\tau_{\Bp}}{\tau_{\Bz}}=1.086\pm0.017$ \cite{pdg}. 
Fig. \ref{fig:combfit} shows the projections of the combined fit results compared 
to the data.
The results 
for the individual mode signal yields and $n_{\mathrm{eff}}$ 
are given in Table~\ref{tab:results}. The significance is computed as 
$\sqrt{2\Delta\log\mathcal{L}}$ where $\Delta\log\mathcal{L}$ is the log 
likelihood difference between the best fit and the null-signal hypothesis.
No significant signal is observed.

\begin{figure}[t]
\includegraphics[width=\linewidth]{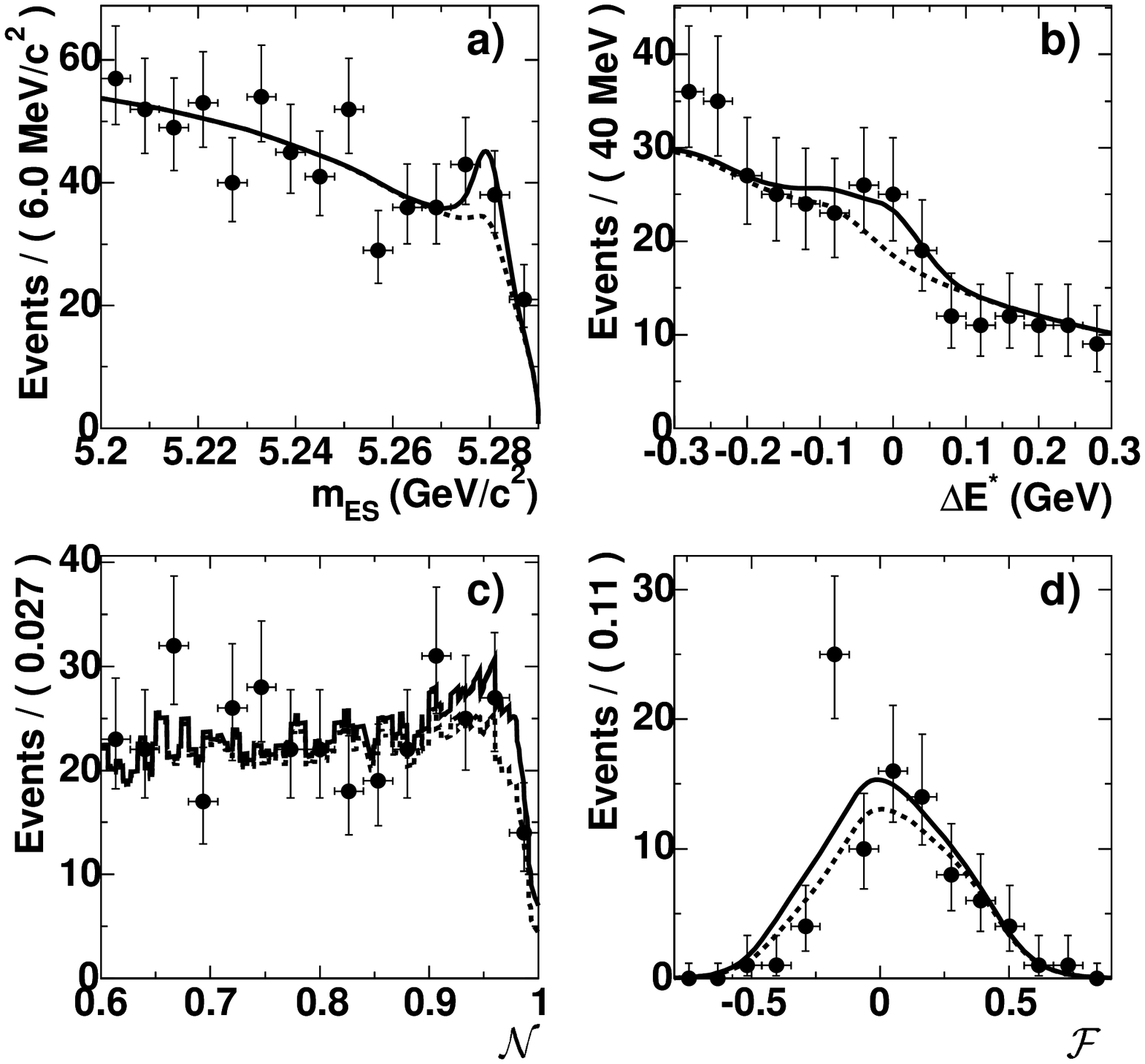}
\vspace{-0.8cm}
\caption{Projections of the combined fit to 
$\brg$ and $\bomg$ in the four discriminating variables: (a) $\mes$, (b) $\DeltaE^{*}$, (c) $\mathcal{N}$, and (d) $\mathcal{F}$. The points are data, 
the solid line is the total PDF and the dashed line is the background 
only PDF. The selections applied, unless the variable is projected, are: 
$5.272<\mes<5.286~\GeVcc$, $-0.10<\de<0.05~\GeV$ and $\mathcal{N}>0.9$; 
the selection efficiencies for signal events are 45\%, 57\%, 70\%, and 44\% for the \mes, \de, $\mathcal{N}$ and $\mathcal{F}$ projections, respectively.}\label{fig:combfit}
\end{figure}

 The most important systematic uncertainties are associated with
the modeling of $\BB$ backgrounds,
the fixed parameters of the PDFs used in the fit, and
the signal reconstruction efficiency.
The first two contribute to the uncertainties on the 
signal yields.

The uncertainty on the peaking \bkg\ background is dominated by the 
\Kpm misidentification rate; the rate is corrected 
by the difference in \Kpm misidentification between data and 
MC samples of $D^{*}$ decays, with the whole correction taken as the 
uncertainty. 
For the  $\Bp\rightarrow\rhop(\piz/\eta)$, $\Bz\rightarrow\rhoz\eta$ and  
$\Bz\rightarrow\omega(\piz/\eta)$ peaking background decays, we vary the 
branching fractions by either one standard deviation from the measured values
or between zero and the measured upper limit if the decay has not been 
observed \cite{charmless,babarrhozpiz}; the value of the 
$\Bz\rightarrow\rhoz\piz$ branching fraction is varied between 
zero and $5.1\times 10^{-6}$ \cite{babarrhozpiz,bellerhozpiz}. The 
uncertainty on the peaking background of each mode is shown in Table~\ref{tab:results}. We find the bias from neglecting the 
$\bkg$ background and combinatorial 
$\BB$ background in the fit to $\bomg$ candidates 
is $1.1^{+1.9}_{-1.1}$ events; the corrected yield is given  
in Table~\ref{tab:results}. To estimate the uncertainty related to the 
extraction of the signal \mes and $\DeltaE^{*}$ PDFs from MC distributions, 
we vary the parameters within their errors. The variation in the fitted 
signal yield is taken as a systematic uncertainty.  
The uncertainty related to the statistics of 
the histogram PDF that describes the continuum $\mathcal{N}$ 
distribution is evaluated by varying the binning and by using  
a fifth-order polynomial as an alternative PDF. 
Several different 
control samples of data and MC events were used to determine alternative 
PDFs for the different hypotheses; none of these resulted in 
a significant change to the fitted signal yield.

The signal efficiency systematic error contains 
uncertainties from tracking, particle identification, 
photon/\piz reconstruction, photon selection and the neural network selection
that are determined as in Ref.~\cite{newkstgam}. 
We determine the effect of correlations among the 
fit variables by using an ensemble of 
MC experiments of parameterized 
continuum background simulations embedded in samples of fully
simulated signal and $\BB$ background events.
No bias is observed within the statistical error on the mean yields from this 
ensemble, which is taken as a multiplicative systematic uncertainty. 
The total multiplicative systematic error values are 
 11\%, 13\% and 10\%  for $\brpg$, $\brzg$ and $\bomg$, respectively. The 
corrected signal efficiencies and their uncertainties are shown in Table~\ref{tab:results}.

In calculating branching fractions, we assume 
${\cal B}(\Upsilon (4S)\! \rightarrow  \BzBzb) = 
{\cal B}(\Upsilon (4S) \! \rightarrow  \BpBm) = 0.5$. 
The $90\%$ C.L. is taken as the largest value of the efficiency-corrected signal yield at which $2\Delta\log\mathcal{L}=1.28^{2}$. We include systematic uncertainties by increasing the efficiency corrected signal yield 
by $1.28$ times its systematic uncertainty. 
Table~\ref{tab:results} shows the resulting 
upper limits on the branching fractions.

Using the measured value of \BR(\bkg) \cite{newkstgam}, 
we calculate a limit of
$\avbr/\BR(\bkg) < 0.029$ at 90\% C.L. 
This limit is used to constrain the ratio of CKM elements
$|V_{td}/V_{ts}|$ by means of the equation \cite{ali2004,alivtdvtstheory}:
\[
\frac{\avbr}{{\cal B}(\bkg)}=
\left| \frac{V_{td}}{V_{ts}} \right|^{2}
\left(\frac{1-m_{\rho}^{2}/M_{B}^{2}}{1-m_{K^{*}}^{2}/M_{B}^{2}}\right)^{3}
\zeta^{2} [1+\Delta R],
\]
where $\zeta$ describes the flavor-SU(3) breaking between $\rho/\omega$ and $K^*$, and
$\Delta R$ accounts for annihilation diagrams. 
Both $\zeta$ and $\Delta R$ must be taken from theory \cite{alivtdvtstheory,ali2004,grinpir}.  
Following \cite{ali2004}, we choose the values 
$\zeta = 0.85 \pm 0.10$, 
and $\Delta R = 0.10 \pm 0.10$, which is the average over the values 
given for the three modes.
We find the limit $|V_{td}|/|V_{ts}| < 0.19$ at 90\% C.L, ignoring 
the theoretical uncertainties. Our upper limit on 
$\Vtd/\Vts$ constrains $\Vtd<0.008$ at 90\% C.L. 
assuming $\Vts=\Vcb$ \cite{pdg}; this  lies within the  
current 90\% confidence interval $0.005 < \Vtd< 0.014$, which is obtained 
from a fit to experimental results on the CKM matrix elements \cite{pdg}. 
Varying the values 
of $\zeta$ and $\Delta R$ within their uncertainties leads to changes in the 
limits by $\pm0.03$ and $\pm0.001$ for $\Vtd/\Vts$ and $\Vtd$, respectively.

In conclusion, we have found no evidence for the exclusive $\bdg$ transitions
$\brg$ and $\bomg$ in 211 million $\FourS\to\BB$ decays studied with the $\babar$ detector.
The $90\%$ C.L. upper limits on the branching fractions and $\Vtd/\Vts$ are
significantly lower than our previous values \cite{oldbabar} and restrict the range indicated by SM predictions \cite{SM,ali2004}.

We are grateful for the excellent luminosity and machine conditions
provided by our \pep2\ colleagues, 
and for the substantial dedicated effort from
the computing organizations that support \babar.
The collaborating institutions wish to thank 
SLAC for its support and kind hospitality. 
This work is supported by
DOE
and NSF (USA),
NSERC (Canada),
IHEP (China),
CEA and
CNRS-IN2P3
(France),
BMBF and DFG
(Germany),
INFN (Italy),
FOM (The Netherlands),
NFR (Norway),
MIST (Russia), and
PPARC (United Kingdom). 
Individuals have received support from CONACyT (Mexico), A.~P.~Sloan Foundation, 
Research Corporation,
and Alexander von Humboldt Foundation.

\end{document}